\documentclass{article}
\usepackage{spconf,amsmath,graphicx,breqn,textcomp,booktabs,array,xcolor}
\usepackage[linesnumbered,ruled,vlined]{algorithm2e}

\newcommand{\iva}{\textsc{iva }}
\newcommand{\hallam}{\textsc{hallam }}
\newcommand{\drintvws}{\textsc{dr intvws }}

\title{DATA AUGMENTATION USING GENERATIVE NETWORKS TO IDENTIFY DEMENTIA}
\name{\begin{tabular}{c}Bahman Mirheidari$^{1}$ \qquad Yilin Pan$^{1}$ \qquad Daniel Blackburn$^{2}$ \qquad Ronan O{\textquotesingle}Malley$^{2}$ \qquad Traci Walker$^{3}$\\ \qquad Annalena Venneri$^{4}$ \qquad Markus Reuber$^{5}$ \qquad Heidi Christensen$^{1}$\end{tabular}}

\address{
\ninept$^{1}$Department of Computer Science, $^2$Sheffield Institute for Translational Neuroscience (SITraN)\\
\ninept$^3$Department of Human Communication Sciences, 
\ninept$^4$Department of Neuroscience, Royal Hallamshire Hospital\\ 
\ninept$^5$Academic Neurology Unit, Royal Hallamshire Hospital\\ 
\ninept\texttt{\{b.mirheidari,yilin.pan, heidi.christensen\}@sheffield.ac.uk}
}

\begin{document}
\topmargin=0mm
\ninept
\maketitle
%
% HC checked
\begin{abstract}
Data limitation is one of the most common issues in training machine learning classifiers for medical application. Due to ethical concerns and data privacy, the number of people that can be recruited to such experiments is generally smaller than the number of participants contributing to non-healthcare datasets. Recent research showed that generative models can be used as an effective approach for data augmentation, which can ultimately help to train more robust classifiers in sparse data domains. A number of studies proved that this data augmentation technique works for image and audio data sets. In this paper, we investigate the application of a similar approach to different types of speech and audio-based features extracted from interactions recorded with our automatic dementia detection system. Using two generative models we show how the generated synthesized samples can improve the performance of a DNN based classifier. The variational autoencoder increased the F-score of a four-way classifier distinguishing the typical patient groups seen in memory clinics from 58\% to around 74\%, a 16\% improvement.
% HC need to add line or two of results (if you have space)
\end{abstract}
\section{Introduction}
\label{sec:intro}   
% HC checked
Dementia is a disorder of cognitive skills affecting memory, everyday functionalities, speech, language and communication abilities. The number of people developing dementia is increasing drastically. It is estimated that there are around 850 thousand people living with dementia in the UK. Dementia is the leading cause of death in the country accounting for over 12 percent of total deaths. The figure has grown by threefold from 2017 to 2005 \cite{dementiastatistics}.~The early diagnosis of dementia is of great clinical importance, and there is a need for an automatic, easy-to-use, low-cost and accurate stratification tool.

% HC checked
Recent studies using the qualitative methodology of conversation analysis (CA) demonstrated that communication problems may be picked up during conversations between patients and neurologists and that this can be used to differentiate between patients with neurodegenerative disorder (ND) and functional memory disorder (FMD; exhibiting problems with memory not caused by dementia)~\cite{Elsey2015,Jones2015}. However, conducting manual CA is expensive and difficult to scale up for routine clinical use.~We have therefore developed a fully automatic system based on analysing a person's speech and language as they speak to an Intelligent Virtual Agent (IVA). The IVA asks a series of memory-probing questions that have been found to be cognitively demanding to answers. These questions are mimicking the style of questions often using during the \textsl{history taking} part of a normal face-to-face consultation. A number of features routed in conversation analysis were extracted and high accuracy levels were achieved when evaluating the system in a real memory clinic on patients with ND and FMD~\cite{mirheidari2017avatar,mirheidari2018,mirheidari2019computational,ronan2019}.~We have recently expanded our data collection to include two more diagnostic classes: healthy controls (HC), and patients with mild cognitive impairment (MCI; a promodal condition to Alzheimer's disease (AD) indicating cognitive decline worse than normal aging but not consistent with an AD diagnosis.) \cite{mirheidari2019computational}.~This changed the task of binary decision for the classifier to a four-way classification, which naturally increased the difficulty due to the large overlap between symptoms (and extracted features) from the HC, FMD and MCI participants. In addition, the amount of data is still limited (in total 60 samples altogether, around 11 hours speech, 3.5 K utterances), which makes it challenging to train a very robust classifier and to apply state-of-the-art deep learning based machine learning techniques successfully.
It is very well-known that to train robust machine learning models, there should be a large number of samples for each class in the training data set; large enough to generalise the model, i.e. predict the classes of unseen samples correctly. However, in the medical domain, the number of people recruited to studies is often limited and the collected datasets are relatively small. Training classifiers with sparse data is therefore a major issue when applying state-of-the-art machine learning in medical applications. Therefore, most research in this field resort to using conventional classifiers rather than the recently introduced deep neural network (DNN) based models.
% HC only describe people as 'authors' if that really is what they are mainly (so, e.g., the authors of this paper ...). I think here you just mean researchers, so have modified it.
% BM DONE

% HC checked
One of the common approaches to increase the number of samples is data augmentation. Data augmentation is widely used in image (\cite{perez2017effectiveness}), speech (\cite{ko2017study}), and text (\cite{semeniuta2017hybrid}) processing to alleviate problems with limited data. The standard augmentation techniques, for instance in image processing, includes rotation, cropping, scaling and transformation of images \cite{bowles2018gan}. There are increasing number of studies applying generative models such as generative adversarial networks (GANs) for data augmentation. For instance in the speech area, GANs used for different tasks such as speech synthesis \cite{kaneko2017generative}, speech recognition \cite{hu2018generative, sheng2018data}, speech emotion recognition \cite{chatziagapi2019data}, speech enhancement \cite{pascual2017segan}, and speaker verification \cite{michelsanti2017conditional}.

% HC done add how we believe this is the first study to explore this; also add that it is not a quarantee that this will work (in the feature domain and for a classification task?)
% BM DONE

% HC checked
In this paper we investigate using three recent generative models to produce synthesized samples of the features extracted from the conversation participants
%between the IVA, patient and accompanying person (if presents). 
Adding the generated features to the original features, we train a new DNN-based classifier to distinguish between the four classes (FMD, ND, MCI and HC). To the best of our knowledge, this is the fist study on direct augmentation of variant statistical features extracted from speech for dementia detection. 
% HC DONE has it be done for other types of dementia detection? Or in speech at all? it sounds like it's only novel because we are doing the combination of conversation and dementia detection?
%This work only is an exploration and feasibility assessment of using the augmentation techniques for improving the performance of classifier.
% BM not been done in dementia detection as far as I am aware
% HC ok - thanks

% HC I moved your new paragraph on the novelty till after you've explained what you're doing.

% HC you haven't described the data when you start talking about IVA, patient etc. I think you can leave those details out here and just refer to this as conversational 

% HC really important point you're making below. Suggest to mention (or hint at) this earlier to really draw out how this is different and worth investigating.
% BM DONE
% HC checked
The majority of generative models introduced for speech, image and text applications are based on CNN or long short-term memory (LSTM), where the order or position of features in each sample is important, and the network learns the context. However, in this paper we aim to use the generative models as an augmentation technique to produce more samples of the features. These features are inherently different in nature to image pixels, speech waveforms or word sequences. Therefore, standard dense layers of neural networks will be used instead of CNNs and LSTMs in the generative models.  
 
\section{Generative models for data augmentation}
\label{sec:background}   

% HC checked
Machine learning models generally fall into two major approaches: discriminative verses generative models.~For the input features, $x$ and the corresponding labels $y$, the discriminative models try to directly make decision boundaries from the features to determine the labels (i.e., directly predict the probability of $y$ given $x$:~$P(y|x)$), while the generative models  focus on feature distribution and generation of features (probability of $x$:~$P(x)$).~In addition to the naive Bayes generative models which have been known for a long time in the machine learning community, there are two recently introduced techniques: GANs and variational autoencoders (VAEs).  

% HC checked
The GAN model, originally introduced by (\cite{goodfellow2014generative}),  consists of two main components: the generator and the discriminator. The generator generates samples (e.g., new images in the case of the MNIST data set\footnote{Large dataset of handwritten digits widely used in the machine learning community \cite{lecun1998mnist}}), while the discriminator authenticates samples, i.e., decides that a sample either comes from the real data set or not. So the task of the discriminator is simply a binary classification. The generator, on the other hand, attempts to deceive the discriminator by creating better samples, as realistic as possible, in order to pass the evaluation of the discriminator (the discriminator gets confused and treats them as the authentic samples). % HC checked
Normally in training GANs, the generator maps randomly made numbers (noise, hidden/latent code) into samples. The synthesised and real samples are both fed to the discriminator, and the discriminator returns a probability indicating authenticity (1:real, 0:fake). The generator model, technically, is built in a reverse network as the discriminator with the opposition loss function. For instance, if the discriminator model is a convolutional network, the generator is the inverse convolutional network. The low resolution of the synthesized sample and the difficulties in stabilising the model are the two main issues of GANs. There have been different improvements to address these issues, including using the Wasserstein distance for the function loss (WGAN \cite{arjovsky2017wasserstein}), conditional GAN (CGAN \cite{mirza2014conditional}) and semi-supervised GANs (SGAN) (forcing the discriminator to produce the labels \cite{odena2016semi}). 

% HC checked
Autoencoder (AE), another generative model, consists of two components: the encoder and the decoder. The encoder encodes the input samples into a compressed representation (latent vectors which are dimensionally reduced version of samples), while the decoder reconstruct the samples from the compressed representations. The aim of AE is to reconstruct the input samples as similar as possible to the real samples. Basically, AE uses an unsupervised training regime to reconstruct the original data. VAEs are the extension to AEs, which normalise the latent vectors. Unlike the conventional AE, VAEs assume Gaussian distribution for the input samples and tries to capture the distribution of the original samples and they are much more similar to GANs than the normal AEs \cite{doersch2016tutorial}.  
% \cite{chatziagapi2019data} used a conditional GAN for generating samples spectrograms for the imbalanced underrepresented emotions in a speech emotion recognition task. Their model, a balancing GAN (BAGAN), uses  convolutional neural network (CNN) layers, initialised by a pre-trained AE for the imbalanced data and fine tuned both for the minority and majority classes. Their augmentation technique improved the unweighted average recall (UAR) of the IEMOCAP data set from 49\% to 53.6\%.

% HC done highlight that these are medical applications but the input data is different 
% BM DONE
% HC checked
Generative models for data augmentation have been used for medical applications, where the data limitation issue is of particular concern. Synthesised samples can help in training more robust classifiers improving the generalisation and reducing the overfitting problem. For instance, GANs have been shown to improve performance in image segmentation tasks such as the computed tomography (CT) cerebrospinal fluid (CSF) and the fluid-attenuated inversion recovery (FLAIR) magnetic resonance (MR) \cite{bowles2018gan}. \cite{frid2018synthetic} reported significant improvement in classification of CT images of liver lesions, when data augmentation was carried out after applying the standard augmentation techniques on the images. Although, these studies were in the medical domain, the data they worked with was very different to our speech data, such as brain and liver images. % HC in the last sentence above: do you mean different to our data or different to each other? It's a bit unclear
% BM DONE

% HC up to here
\section{Experimental setup}
\label{sec:experimental_setup}
% HC checked
The data was collected using the IVA during summers of~2016, 2017, 2018 and 2019 at the Department of Neurology, University of Sheffield, UK based at the Royal Hallamshire Hospital. Of the total number of 93 participants, 60 were chosen for the study (the rest were found to not have memory problems, however we made use of that data for training the speaker diarisation and the ASR).~Table~\ref{tab:demographic_info} shows the demographic information of the participants in the study.~Comparing to the previous experiment (\cite{mirheidari2019computational}), in this study we use a balanced number of conversations for each class of our four groups (i.e. 15 FMD, 15 ND, 15 MCI and 15 ND).  

\begin{table}[h]

% HC is clear what 'education' means in the table? Assume it's years in education? Is there space to add this in caption?
% HC checked
\small
\caption{Demographic information of the participants (15 in each group).~FMD:~Functional Memory Disorder, ND:~Neuro-degenerative Disorder, MCI:~Mild Cognitive Impairment, HC:~Healthy Control. }
\label{tab:demographic_info}
% \vspace{2mm}
\centering{ 
\begin{tabular}{c | c |c  |c} 
\hline
\multicolumn{1}{c|}{\textbf{Class}} &  
\multicolumn{1}{c|}{\textbf{Age}} &
\multicolumn{1}{c|}{\textbf{Education (Years)}} &
\multicolumn{1}{c}{\textbf{Male}}\\
\hline
FMD &  54.9 (+/- 4.1) & 16.4 (+/- 0.6) & 40.0\%   \\
\hline
ND &   67.8 (+/- 4.2)  & 18.0 (+/- 1.6)& 66.7\%   \\
\hline
MCI &  63.0 (+/- 4.3) & 17.3 (+/- 1.1) & 66.7\%   \\
\hline
HC &  69.5 (+/- 4.0)  & 18.1 (+/- 1.0)& 40.0\%   \\
\hline
% \hline
% All & 60 & 65.3 (+/- 9.62) & 56.4 (+/- 6.22) & 44.3\%   \\
% \hline
\end{tabular}
}
\end{table} 

% HC checked
 Table~\ref{tab:datasets_info} shows the information of the two datasets used:~\drintvws (295 doctor-patient interviews) and \iva (93 IVA-patient recordings).~The \drintvws data set was only used for training the i-vector based diarisation module (the CALL\_HOME recipe \cite{prince2007probabilistic}) and the Bidirectional Long Short Term Memory/Time-Delay Neural Network (BLSTM)-TDNN based ASR using the Kaldi toolkit \cite{Povey_ASRU2011}.
The $10$ fold cross validation approach was used for training the diarisation and the ASRs. The diarisation error rate (\textbf{DER}) was \textbf{26.2\%}, and the word error rate (\textbf{WER}) was \textbf{38.2\%}. 

% HC there's a trick to making letters inside equations look nice in Latex - can't remember what it is though .. mighe be \textrm{}
% HC checked
\subsection{Extended features}
In addition to the initial features (78 including CA-inspired, only-acoustic, only-lexical, word vector and verbal fluency) introduced in \cite{mirheidari2019computational}), 104 MFCC acoustic features were extracted. Then the min, max, average and standard deviation were applied ($4 (\textrm{func}.) \times 13 (\textrm{MFCC}) \times 2 (\textrm{speakers}) = 104$). The initial acoustic-only and lexical-only features included the average as the only statistic function, therefore the other remaining three statistic features were also applied on the features, resulting in additional 72 acoustic-only ($3 (\textrm{func.}) \times 24 (\textrm{acoustic-only}) = 72$) and 72 lexical-only features ($3 (\textrm{func.}) \times 24 (\textrm{lexical-only}) = 72$). This resulted in a total number of 324 features.

% HC checked
\begin{table}[h]

\caption{\label{tab:datasets_info} Datasets used for training the ASRs, including Len.:the total length in hours/mins, Utts.:number of utterances, Spks.:number of speakers, and Avg. Utts.:Average utterance length in seconds.}
% \vspace{2mm}
\centering{
\begin{tabular}{p{2cm}  |p{1.3cm}  | p{0.4cm}|p{0.4cm}  | p{0.4cm}}
\specialrule{0.8pt}{0.8pt}{0.8pt} 
\multicolumn{1}{c|}{\textbf{Dataset(No)}} &  
\multicolumn{1}{c|}{\textbf{Len.}} & 
\multicolumn{1}{c|}{\textbf{Utts. }}&  
\multicolumn{1}{c|}{\textbf{Spks.}} & 
\multicolumn{1}{c}{\textbf{Avg Utts.}}\\ 
\specialrule{0.8pt}{0.8pt}{0.8pt}
% \mbox{IMDB(50K)} &-&-&-&- \\
% \hline  
% \mbox{Dem (551)} &8h 34m&6737&293&4.6s \\ 
% \hline
%\mbox{\hallam+Seizure(295)}&64h 21m&39184&736&5.9s\\
\mbox{Dr intvws (295)}&64h 21m&39184&736&5.9s\\
\hline
\mbox{\iva{} (93)}&17h 18m&5637&103&11.05s\\   
\specialrule{0.8pt}{0.8pt}{0.8pt}
\end{tabular}
} 
\end{table}   

% HC checked
\subsection{Details of the generative models}
For training the generative models we used the Keras python library (\cite{chollet2015keras}) back-ended by Tensorflow(\cite{tensorflow2015}). Three candidate generative models were selected: CGAN, VAE and VAE combined with SGAN. This is similar to the AE-GAN introduced by \cite{makhzani2015adversarial}, but the CNN layers were replaced with  dense layers, and the AE with VAE; we refer to this model as VAE-SGAN. The encoder and decoder parts of the VAE-SGAN were similar to the encoder and decoder of the VAE. In addition to the normal dense layers, and to reduce overfitting, layers of BatchNormalization, LeakyReLU, and Dropout were used in between the layers. The Adam optimizer was used for training, as well as a two layer standard DNN classifier which is used separately to evaluate the synthesized samples.  

% HC fixed
Algorithm \ref{alg:gen} shows how we use the generative models to make synthesized samples and add them to the training set. We can repeat this $N$ times and keep the results for both the test and evaluation (eval) sets separately. In order to see how well the reconstructed samples do, a DNN based classifier is used and the F-score is calculated based on its performance on both the test and eval sets.
% HC HC where are these results? You should either comment on them here, refer to a table or let the reader know that these results will be discussed in the next section

% HC is this a fairly standard way of reporting this type of info? 
% ***I think so, but to save space I can summarise them in words
% \begin{table}[h]
% \caption{\label{tab:layers} Layers of the discriminator, encoder and, decoder of the generative models.}
% % \vspace{2mm}
% \centering{
% \begin{tabular}{p{2cm}  | p{4cm} }
% \specialrule{0.8pt}{0.8pt}{0.8pt} 
% \multicolumn{1}{c|}{\textbf{Dis. layers}}&  
% \multicolumn{1}{c}{\textbf{Configuration}}\\ 
% \specialrule{0.8pt}{0.8pt}{0.8pt}  
%  Dense & 1024 * lat-dim \\  
%  LeakyReLU &  alpha=0.2\\
%  Dense & 512 *  \\  
%  LeakyReLU &  alpha=0.2\\
%   Dense & 1 * , act:sigmoid \\   
% \specialrule{0.8pt}{0.8pt}{0.8pt}    
% \multicolumn{1}{c|}{\textbf{Enc. layers}}&  
% \multicolumn{1}{c}{\textbf{Configuration}}\\ 
% \specialrule{0.8pt}{0.8pt}{0.8pt}  
%  Dense & 1024 * feature-shape \\  
%  LeakyReLU &  alpha=0.2\\
%  Dense & 512 * feature-shape \\  
%  LeakyReLU &  alpha=0.2\\
%   Dense & 1024 * lat-dim \\  
%  Lambda &  \\
% \specialrule{0.8pt}{0.8pt}{0.8pt}   
% \multicolumn{1}{c|}{\textbf{Dec. layers}}&  
% \multicolumn{1}{c}{\textbf{Configuration}}\\ 
% \specialrule{0.8pt}{0.8pt}{0.8pt}  
%  Dense & 512 * lat-dim \\  
%  LeakyReLU &  alpha=0.2\\
%  Dense & 1024 *   \\  
%  LeakyReLU &  alpha=0.2\\
%   Dense &  feature-shape, act:tanh \\   
% \specialrule{0.8pt}{0.8pt}{0.8pt}   
% \end{tabular}
% } 
% \end{table}  

\begin{algorithm} 
\SetAlgoLined
\KwResult{Best scores and reconstruction numbers for the eval and test data:~$Score_{eval}, Score_{test}$}
 \textbf{Input:}~train, eval and test data:~$X_{train}, Y_{train}, X_{eval}, Y_{eval}, X_{test}, Y_{test}$\;
 $reconX=X_{train}$;$reconY=Y_{train}$\;
 $Score_{eval}=(0,0);$ $Score_{test}=(0,0)$\;
 \For{$recon=1, 2, ..., N$}{
  Train a generative model, M(Enc,Dec,Dis)\
  with $X_{train}, Y_{train}$\;
  $lat=Enc(X_{train})$\;
  $X^\prime=Dec(lat)$\;
  $reconX, reconY = reconX + X^\prime,reconY + Y_{train}$\;
  $lat2=Enc(X_{eval})$\;
  $X2^\prime=Dec(lat2)$\;
  $reconX2, reconY2 = reconX + X2^\prime,reconY2 + Y_{eval}$\;
  Train a DNN-based model, DM \
  with $reconX, reconY$ tuned by $reconX2, reconY2$\; 
  $S_{eval}=DM.score(X_{eval}, Y_{eval})$\;
  $S_{test}=DM.score(X_{test}, Y_{test})$\;
  \If{$S_{eval}\geq Score_{eval}[0]$}{
  $Score_{eval}=(S_{eval},rec)$\;
  } 
  \If{$S_{test}\geq Score_{test}[0]$}{
  $Score_{test}=(S_{test},rec)$\;
  } 
 } 
 return $Score_{test}, Score_{test}$\;
 \caption{ \label{alg:gen}Reconstructed samples from a generative model.}
\end{algorithm} 

% HC this section - section headings? none of them seems to be classification experiments according to headings, for example?
\section{Results}
\label{sec:Results}
% HC checked
This section compares the performance on a normal classifier baseline (logistic regression - LR) and a DNN-based trained using the adding the synthesized samples.
% HC  or add brief intro sentence to remind the reader what you are trying to investigate and how? The two main systems you're comparing (with and without augmentation)
% HC HC it's slightly confusing that 

\subsection{Normal classifier}
% HC checked
Using the LR classifier and the five fold cross validation approach the precision, recall and F-score of the classifier were calculated first on the original 78 features and then on the 324 features (original+extended features). The columns with majority of zero values were omitted from the feature sets (we call them non-zeros (NZ)). We observed that using the NZ can result in a better performance for the recursive feature elimination (RFE, a standard approach for feature selection) (\cite{scikit-learn}). Based on the five fold cross-validation, in each fold out of the total 60 samples, 40 were used in train set, 8 for evaluation and 12 for test. Table \ref{tab:LR_HAL} shows the details of the performance of the classifier in terms of precision, recall and F-score for the original 78 features, the NZ original features, the top 13 original features selected by RFE, all features (original+extended), the NZ for all features, and the top 68 all features selected by RFE.  It can be seen, that the NZ features from the original set can achieve around 40\% F-score (3.5\% increase), which then can be improved further by RFE up to 59\%. However, using all features together resulted in a better performance than the original features (F-score of 45.6\% compared to the 36.6\% F-score). Applying RFE (68 top features) this was further improved to an F-score of 64\% (F-score of fold 5 was 58.3\%, the closest F-scores to the average).
% HC checked
On the last row of the table, the results for the average fold (number 5) is shown. We will refer to this fold as (\hallam (F5)). This fold will be used in the following experiments as a fixed train/test partition.

% \begin{table}
% \caption{\label{tab:LR_HAL} Precision (Pr), recall (Rc) and F-scorer (Fs) of the Logistic Regression classifier trained using the 46 and 152 features extracted from the 60 conversations with 5-fold cross validation.}
% % \vspace{2mm}
% \centering{
% \begin{tabular}{p{1cm}  |p{1cm} |p{1.1cm}  | p{1.1cm}|p{1.1cm} }
% \specialrule{0.8pt}{0.8pt}{0.8pt} 
% \multicolumn{1}{c|}{\textbf{Feat. No.}} & 
% \multicolumn{1}{c|}{\textbf{Fold No.}} &  
% \multicolumn{1}{c|}{\textbf{Pr \%}} & 
% \multicolumn{1}{c|}{\textbf{Rc \%}}&  
% \multicolumn{1}{c}{\textbf{Fs \%}}\\ 
% \specialrule{0.8pt}{0.8pt}{0.8pt}  
% 46 & 1 &45.8&41.7&41.1 \\ 
% \textbf{46}  & \textbf{2*} &\textbf{29.6}&\textbf{41.7}&\textbf{34.2} \\
% 46  & 3 &41.7 & 41.7 & 41.0 \\
% 46  & 4 &17.7 &33.3 & 22.0 \\
% 46  & 5 & 52.1 & 41.7 & 42.2 \\
% 46  & - & 37.4(13.7) & 40.0(3.8) & 36.1(8.5) \\
% \specialrule{0.8pt}{0.8pt}{0.8pt}
% 152 & 1 &56.7&58.3&55.4\\
% \textbf{152} & \textbf{2*} &\textbf{47.9}&\textbf{41.7}&\textbf{43.8}\\
% 152 & 3 &64.6&58.3&56.6\\
% 152 & 4 &27.1&25.0&25.5\\
% 152 & 5&35.4&33.3&33.8\\
% 152 & - & 46.3(15.3) & 43.3(14.9) & 43.0(13.5) \\
% \specialrule{0.8pt}{0.8pt}{0.8pt}
% \end{tabular}
% } 
% \end{table}  

% HC checked
\begin{table}[h]
\caption{\label{tab:LR_HAL} Precision (Pr), recall (Rc) and F-scores (Fs) of the Logistic Regression classifier trained using different sets of features extracted from the 60 conversations with 5-fold cross validation. NZ:~Non-zero features.
%Note: Before RFE, we removed the zero features (features which are zeros in more than a half samples in each class. 
F5: Fold 5, the fold close to the average.)}
% \vspace{2mm}
\centering{
\begin{tabular}{p{2.7cm}  |p{0.5cm} |p{0.7cm}  | p{0.7cm}|p{0.7cm} }
\specialrule{0.8pt}{0.8pt}{0.8pt} 
\multicolumn{1}{c|}{\textbf{Feature set}} & 
\multicolumn{1}{c|}{\textbf{Feat. No.}} &  
\multicolumn{1}{c|}{\textbf{Pr \%}} & 
\multicolumn{1}{c|}{\textbf{Rc \%}}&  
\multicolumn{1}{c}{\textbf{Fs \%}}\\ 
\specialrule{0.8pt}{0.8pt}{0.8pt}  
Original & 78 & 36.7 & 40.0 & 36.6  \\ 
Original (NZ) & 64 & 41.5 & 43.3 & 40.1  \\
Original (RFE) & 13 & 60.5 & 60.0 & 59.1  \\ 
\specialrule{0.8pt}{0.8pt}{0.8pt}
ALL & 324 & 45.9 & 46.6 & 45.6 \\ 
ALL (NZ) & 261 & 45.1  & 45.0 & 44.4  \\ 
ALL (RFE) & 68 & 64.5  & \textbf{65.0} & \textbf{64.0}  \\ 
\specialrule{0.8pt}{0.8pt}{0.8pt} 
ALL (F5) (RFE) & 68 & \textbf{68.3} & 58.3 & 58.3  \\ 
\specialrule{0.8pt}{0.8pt}{0.8pt}
\end{tabular}
} 
\end{table}  

% RFE=13 5sec wav
% 2019-10-19 17:54:55.401638 - n-lr point 0 - precision:50.85 recall:50.00 fscore:49.76
% fold 5:2019-10-19 17:54:55.383259 - n-lr - fold 5 - best point 0 - precision(v):29.17 recall(v):50.00 fscore(v):36.67 - precision:58.75 recall:50.00 fscore:48.39

%2.5, RFE=7 2019-10-19 18:13:59.643993 - n-lr point 0 - precision:57.78 recall:56.67 fscore:55.57
% 2019-10-19 18:13:59.622106 - n-lr - fold 3 - best point 0 - precision(v):75.00 recall(v):62.50 fscore(v):62.50 - precision:56.67 recall:58.33 fscore:55.42

% 2019-10-18 12:46:28.836952 - n-lr point 0 - precision:17.73 recall:23.33 fscore:20.08
% e. one sec
% f. two secs
% g. three secs

% 2.5sec 2019-10-19 18:30:25.010126 - n-lr point 0 - precision:31.82 recall:31.67 fscore:31.56  2019-10-19 18:30:24.989604 - n-lr - fold 4 - best point 0 - precision(v):16.67 recall(v):25.00 fscore(v):20.00 - precision:30.83 recall:33.33 fscore:30.83

% HC Explain below whey you're first running experiments on this dataset (to try out the new approach/different network) on a well-known dataset; and what you will be exploring: the augmentation factor.
% HC is 'Feat. no' a good name for this? It's the number of times you're 
\subsection{Generative models on MNIST}
% HC section
% HC checked
Before we start using the augmentation techniques on our dataset, we demonstrate the approach on a widely used dataset, MNIST. This experiment will show how the technique generally works on a standard data set. 
%This experiment will show how the technique generally works on a standard data set. 
MNIST contains 60000 train and 10000 test hand written digit images (in 10 classes, each 28 by 28 pixels). Generative models have been shown to work well for tasks involving images, speech and text where there are sequences of features in which the neighbouring features may be co-related to each other. As mentioned before, we removed all CNN or LSTM layers (which capture context information very well). So as expected, this reduces the performance of the generative model significantly. % HC is this shown in a table? If not, you should put the numbers in here. Otherwise, make sure to refer to the table
From the train set of MNIST, 1500 samples (150 for each digit) were selected. 10 percent was chosen as the eval set (150 samples) and 90 percent (1350 samples) as the train set (we refer to this as MNIST-1500). The standard 10K samples of MNIST was used as the test in our experiments.
% HC I put 'related' instead of 'co-related' above. Happy to change back if you definitely feel it should be co-related.

% \begin{figure}[htb] 
% \begin{minipage}[b]{0.5\linewidth}
%   \centering
%   \centerline{\includegraphics[width=6cm]{Encoder.png}}
% %  \vspace{2.0cm}
%   \centerline{(a) Variational Encoder}\medskip
% \end{minipage}
% %
% \begin{minipage}[b]{0.55\linewidth}
%   \centering
%   \centerline{\includegraphics[width=6cm]{Decoder.png}}
% %  \vspace{1.5cm}
%   \centerline{(b) Decoder}\medskip
% \end{minipage} 
% %
% \caption{Variational Encoder's architecture.}
% \label{fig:res}
% %
% \end{figure}

\begin{figure}[h]
\centering 
\includegraphics[width=0.85\linewidth,height=0.6\linewidth]{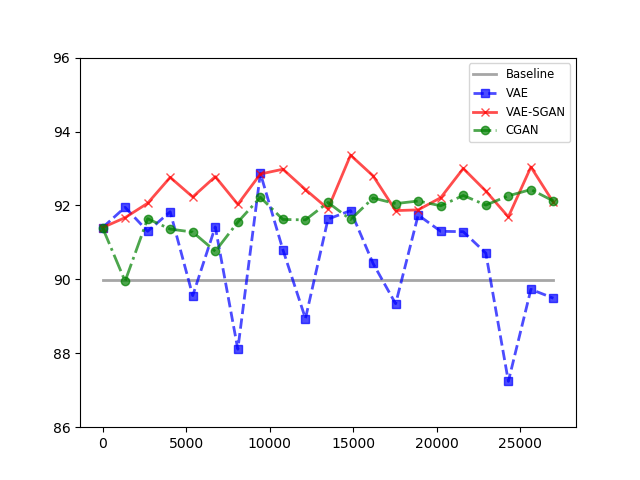}
\caption{\it{F-scores (\%) of the DNN classifier for the MNIST-1500 data for different numbers of reconstructed samples tested on the standard 10K samples of the test set (number of reconstructions:~20).}}
\label{fig:mnist_fscores}
\end{figure}

% HC checked
The F-score, when training the normal LR classifier (baseline) on the MNIST-1500 subset, was around 90\%. The algorithm for augmentation was applied on the dataset using the three generative models. Figure~\ref{fig:mnist_fscores} shows the F-scores gained using the algorithm over 20 times reconstructions (up to 27000 additional samples). As can be seen, all three generative models (CGAN, VAE and VAE-SGAN) improved the F-score up to around 93\%. The improvement seen is not steady though, and the results fluctuate, however on average, VAE-SGAN performed slightly better than CGAN, while VAE was not as good as the other two and had the highest fluctuation.

% HC checked

% HC checked
\subsection{Generative models on \hallam (F5)}
\begin{figure}[h]
\centering
\includegraphics[width=0.85\linewidth,height=0.6\linewidth]{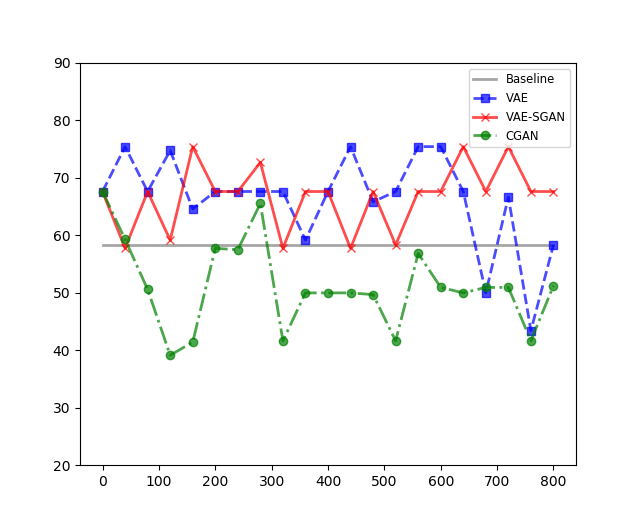}
\caption{\it{F-scores (\%) of the DNN classifier for the \hallam(F5) data for different numbers of reconstructed samples tested on 12 samples of the test set (number of reconstructions:~25).}}
\label{fig:hal_fscores}
\end{figure}

The algorithm was repeated for the \hallam (F5) data set. The baseline classifier F-score was around 58\%. Figure \ref{fig:hal_fscores} shows the F-scores when applying the three generative models. As the number of reconstructed features increased (up to 2000), the F-scores of the VAE-SGAN varies between 60\% to around 75\%, VAE fluctuated between 40\% to 74\%, and CGAN between 40\% to 70\%. Compared to MNIST-1500, these fluctuations were much higher. VAE-SGAN, however, performed better.

% HC checked
% HC is it F5 or F5?

\subsection{Optimum number of reconstructions}
% HC checked
Based on the previous two figures, finding the optimum number of reconstructions is challenging. One approach is to use the eval set to find the best F-score, although naturally, this might not be the best for the test set. The high fluctuation, especially in \hallam (F5) might indicate that not all of the reconstructed samples are useful for the classification. Therefore, we modified the Algorithm \ref{alg:gen}, to first check the quality of the reconstructed features, and only if they are good enough, are they then added to the train set. We used a similarity measure (normalised pair-wise distance) between the two individual features in the feature set (similar to the pixel-based similarity in image processing). We observed, that we can get better results if in each iteration of the algorithm, we check whether the similarity improves or not between the reconstructed features and the original features in the train set.
% Table \ref{tab:best_test_best} shows the global optimum number of samples and the F-scores for the two data sets. Since the test sets are totally separate, we can not get these optimum numbers only using the eval sets. 
% HC it's not clear to me how you have found this optimum number that you have in Table 4. Can you add a sentence or so to explain in which that's the optimum number?
% HC checked
Table~\ref{tab:results_refined} shows the best F-scores gained for MNIST and \hallam (F5) after modifying the algorithm. Using VAE, MNIST achieved a 90.3\% F-score, while the VAE-SGAN and CGAN achieved 92.3\%. For \hallam (F5), VAE performed the best with a 75.4\% F-score. CGAN gained 74.8\%. Comparing to the baseline of 58.3\%, all three generative models saw improvements with VAE performing the best at 17.1\%, and CGAN and VAE-SGAN following with 16.5\% 9.3\%, respectively. 

% HC checked
% HC move the % nto the table; right adjust all numbers
\begin{table} 
\caption{\label{tab:results_refined} {F-scores (Fs) and the number of reconstructed samples after modifying the algorithm (see text for details).}}
% \vspace{2mm}
\centering{
\begin{tabular}{p{1.8cm} |p{1.6cm}  |p{0.6cm} |p{0.6cm}  }
\specialrule{0.8pt}{0.8pt}{0.8pt} 
\multicolumn{1}{c|}{\textbf{Data set}} &
\multicolumn{1}{c|}{\textbf{Model}} &
\multicolumn{1}{c|}{\textbf{Train no.}} &   
\multicolumn{1}{c}{\textbf{Fs}}\\
\specialrule{0.8pt}{0.8pt}{0.8pt}   
MNIST-1500 &VAE  & 20250  & 90.3\%\\ 
MNIST-1500 &CGAN  & 2700  & \textbf{92.3\%}  \\
MNIST-1500 &VAE-SGAN  & 4050  & \textbf{92.3\%} \\ 
\specialrule{0.8pt}{0.8pt}{0.8pt}  
\hallam(F5)  & VAE & 160  & \textbf{75.4\%} \\ 
\hallam(F5)  & CGAN & 200 & 74.8\%  \\ 
\hallam(F5)  & VAE-SGAN & 120 &  67.6\% \\ 
\specialrule{0.8pt}{0.8pt}{0.8pt} 
\end{tabular}
} 
\end{table}

\section{Conclusions}
We introduced the concept of using data augmentation on features extracted from a person's speech and language using the recent generative models CGAN, VAE and VAE-SGAN. Finding the optimum number of reconstructed samples is the challenging part of this technique, although the evaluation set can help us to find a local optimum number. For the two tasks using Algorithm \ref{alg:gen} and the modification, each generative model performed well, if slightly differently. However more work is needed to investigate the use of generative models. Reported experiments were carried out on a representative fold; future work will expand this to all fold as well as exploring the effect of different generative models with more reconstructions. 
%Due to the computation and time restrictions we only performed the experiment on one  of our five fold data, perhaps we can repeat this for all the folds. As a further work we can explore more generative models with more reconstructions.

% HC we should also add the MRC CinC acknowledgements - have we got an example of this?
% ******* where I can find it?
\section{Acknowledgements}
This research has been partly supported under the European Union's H2020 Marie Skłodowska-Curie programme TAPAS (Training Network for Pathological Speech processing; Grant Agreement No. 766287)

\clearpage
\newpage
\bibliographystyle{IEEEbib}
\bibliography{refs}

\begin{thebibliography}{10}

\bibitem{dementiastatistics}
{Dementia Statistics},
\newblock ``Deaths due to dementia,'' 2018,
\newblock Accessed on October 12, 2019.

\bibitem{Elsey2015}
C.~Elsey, P.~Drew, D.~Jones, D.~Blackburn, S.~Wakefield, K.~Harkness,
  A.~Venneri, and M.~Reuber,
\newblock ``{Towards diagnostic conversational profiles of patients presenting
  with dementia or functional memory disorders to memory clinics},''
\newblock {\em Patient Education and Counseling}, vol. 98, pp. 1071--1077,
  2015.

\bibitem{Jones2015}
D.~Jones, P.~Drew, C.~Elsey, D.~Blackburn, S.~Wakefield, K.~Harkness, and
  M.~Reuber,
\newblock ``{Conversational assessment in memory clinic encounters:
  interactional profiling for differentiating dementia from functional memory
  disorders},''
\newblock {\em Aging {\&} Mental Health}, vol. 7863, pp. 1--10, 2015.

\bibitem{mirheidari2017avatar}
B.~Mirheidari, D.~Blackburn, K.~Harkness, T.~Walker, A.~Venneri, M.~Reuber, and
  H.~Christensen,
\newblock ``An avatar-based system for identifying individuals likely to
  develop dementia,''
\newblock {\em Proc.\ Interspeech}, pp. 3147--3151, 2017.

\bibitem{mirheidari2018}
B.~Mirheidari, D.~Blackburn, A.~Venneri, M.~Reuber, T.~Walker, and
  H.~Christensen,
\newblock ``Detecting signs of dementia using word vector representations,''
\newblock in {\em Proc. Interspeech}. ISCA, 2018.

\bibitem{mirheidari2019computational}
B.~Mirheidari, D.~Blackburn, R.~O’Malley, T.~Walker, A.~Venneri, M.~Reuber,
  and H.~Christensen,
\newblock ``Computational cognitive assessment: Investigating the use of an
  intelligent virtual agent for the detection of early signs of dementia,''
\newblock in {\em ICASSP 2019-2019 IEEE International Conference on Acoustics,
  Speech and Signal Processing (ICASSP)}. IEEE, 2019, pp. 2732--2736.

\bibitem{ronan2019}
R~O’Malley, B.~Mirheidari, A.~Harkness, K.and~Venneri, M.~Reuber, T.~Walker,
  H.~Christensen, and D.~Blackburn,
\newblock ``A fully automated cognitive screening tool based on assessment of
  speech and language,''
\newblock {\em Journal of Neurology, Neurosurgery \& Psychiatry}, 2019,
\newblock In preparation.

\bibitem{perez2017effectiveness}
L.~Perez and J.~Wang,
\newblock ``The effectiveness of data augmentation in image classification
  using deep learning,''
\newblock {\em arXiv preprint arXiv:1712.04621}, 2017.

\bibitem{ko2017study}
T.~Ko, V.~Peddinti, D.~Povey, M.~L Seltzer, and S.~Khudanpur,
\newblock ``A study on data augmentation of reverberant speech for robust
  speech recognition,''
\newblock in {\em 2017 IEEE International Conference on Acoustics, Speech and
  Signal Processing (ICASSP)}. IEEE, 2017, pp. 5220--5224.

\bibitem{semeniuta2017hybrid}
S.~Semeniuta, A.~Severyn, and E.~Barth,
\newblock ``A hybrid convolutional variational autoencoder for text
  generation,''
\newblock {\em arXiv preprint arXiv:1702.02390}, 2017.

\bibitem{bowles2018gan}
C.~Bowles, L.~Chen, R.~Guerrero, P.~Bentley, R.~Gunn, A.~Hammers, Maria~V.
  Dickie, D. Alexander~H., J.~Wardlaw, and D.~Rueckert,
\newblock ``Gan augmentation: augmenting training data using generative
  adversarial networks,''
\newblock {\em arXiv preprint arXiv:1810.10863}, 2018.

\bibitem{kaneko2017generative}
T.~Kaneko, H.~Kameoka, N.~Hojo, Y.~Ijima, K.~Hiramatsu, and K.~Kashino,
\newblock ``Generative adversarial network-based postfilter for statistical
  parametric speech synthesis,''
\newblock in {\em 2017 IEEE International Conference on Acoustics, Speech and
  Signal Processing (ICASSP)}. IEEE, 2017, pp. 4910--4914.

\bibitem{hu2018generative}
Hu~Hu, Tian Tan, and Yanmin Qian,
\newblock ``Generative adversarial networks based data augmentation for noise
  robust speech recognition,''
\newblock in {\em 2018 IEEE International Conference on Acoustics, Speech and
  Signal Processing (ICASSP)}. IEEE, 2018, pp. 5044--5048.

\bibitem{sheng2018data}
P.~Sheng, Z.~Yang, H.~Hu, T.~Tan, and Y.~Qian,
\newblock ``Data augmentation using conditional generative adversarial networks
  for robust speech recognition,''
\newblock in {\em 2018 11th International Symposium on Chinese Spoken Language
  Processing (ISCSLP)}. IEEE, 2018, pp. 121--125.

\bibitem{chatziagapi2019data}
A.~Chatziagapi, G.~Paraskevopoulos, D.~Sgouropoulos, G.~Pantazopoulos,
  M.~Nikandrou, T.~Giannakopoulos, A.~Katsamanis, A.~Potamianos, and
  S.~Narayanan,
\newblock ``Data augmentation using gans for speech emotion recognition,''
\newblock {\em Proc. Interspeech 2019}, pp. 171--175, 2019.

\bibitem{pascual2017segan}
S.~Pascual, A.~Bonafonte, and J.~Serra,
\newblock ``Segan: Speech enhancement generative adversarial network,''
\newblock {\em arXiv preprint arXiv:1703.09452}, 2017.

\bibitem{michelsanti2017conditional}
D.~Michelsanti and Z.~Tan,
\newblock ``Conditional generative adversarial networks for speech enhancement
  and noise-robust speaker verification,''
\newblock {\em arXiv preprint arXiv:1709.01703}, 2017.

\bibitem{goodfellow2014generative}
I.~Goodfellow, J.~Pouget-Abadie, M.~Mirza, B.~Xu, D.~Warde-Farley, S.~Ozair,
  A.~Courville, and Y.~Bengio,
\newblock ``Generative adversarial nets,''
\newblock in {\em Advances in neural information processing systems}, 2014, pp.
  2672--2680.

\bibitem{lecun1998mnist}
Y.~LeCun,
\newblock ``The mnist database of handwritten digits,''
\newblock {\em http://yann. lecun. com/exdb/mnist/}.

\bibitem{arjovsky2017wasserstein}
M.~Arjovsky, S.~Chintala, and L.~Bottou,
\newblock ``Wasserstein gan,''
\newblock {\em arXiv preprint arXiv:1701.07875}, 2017.

\bibitem{mirza2014conditional}
M.~Mirza and S.~Osindero,
\newblock ``Conditional generative adversarial nets,''
\newblock {\em arXiv preprint arXiv:1411.1784}, 2014.

\bibitem{odena2016semi}
A.~Odena,
\newblock ``Semi-supervised learning with generative adversarial networks,''
\newblock {\em arXiv preprint arXiv:1606.01583}, 2016.

\bibitem{doersch2016tutorial}
C.~Doersch,
\newblock ``Tutorial on variational autoencoders,''
\newblock {\em arXiv preprint arXiv:1606.05908}, 2016.

\bibitem{frid2018synthetic}
M.~Frid-Adar, E.~Klang, M.~Amitai, J.~Goldberger, and H.~Greenspan,
\newblock ``Synthetic data augmentation using gan for improved liver lesion
  classification,''
\newblock in {\em 2018 IEEE 15th international symposium on biomedical imaging
  (ISBI 2018)}. IEEE, 2018, pp. 289--293.

\bibitem{prince2007probabilistic}
S.~Prince and J.~Elder,
\newblock ``Probabilistic linear discriminant analysis for inferences about
  identity,''
\newblock in {\em Computer Vision. IEEE 11th International Conference}, 2007,
  pp. 1--8.

\bibitem{Povey_ASRU2011}
D.~Povey, A.~Ghoshal, G.~Boulianne, L.~Burget, O.~Glembek, N.~Goel,
  M.~Hannemann, P.~Motlicek, Y.~Qian, P.~Schwarz, J.~Silovsky, G.~Stemmer, and
  K.~Vesely,
\newblock ``The kaldi speech recognition toolkit,''
\newblock in {\em IEEE 2011 Workshop on Automatic Speech Recognition and
  Understanding}, 2011.

\bibitem{chollet2015keras}
F.~Chollet et~al.,
\newblock ``Keras: Deep learning library for theano and tensorflow,''
\newblock {\em URL: https://keras. io/k}, vol. 7, pp. 8, 2015.

\bibitem{tensorflow2015}
M.~Abadi et~al.,
\newblock ``{TensorFlow}: Large-scale machine learning on heterogeneous
  systems,'' 2015,
\newblock Software available from tensorflow.org.

\bibitem{makhzani2015adversarial}
A.~Makhzani, J.~Shlens, N.~Jaitly, I.~Goodfellow, and B.~Frey,
\newblock ``Adversarial autoencoders,''
\newblock {\em arXiv preprint arXiv:1511.05644}, 2015.

\bibitem{scikit-learn}
F.~Pedregosa, G.~Varoquaux, A.~Gramfort, V.~Michel, B.~Thirion, O.~Grisel,
  M.~Blondel, P.~Prettenhofer, R.~Weiss, V.~Dubourg, J.~Vanderplas, A.~Passos,
  D.~Cournapeau, M.~Brucher, M.~Perrot, and E.~Duchesnay,
\newblock ``Scikit-learn: Machine learning in {P}ython,''
\newblock {\em Journal of Machine Learning Research}, vol. 12, pp. 2825--2830,
  2011.

\end{thebibliography}

\end{document}